\documentclass[preprint,showpacs,amsmath,amssymb,floatfix,pre]{revtex4}
\usepackage{graphicx}
\begin{document}
\title{Spin-$1/2$ anisotropic Heisenberg antiferromagnet with Dzyaloshinskii-Moriya interaction 
via mean-field approximation} 

\author{Walter E. F. Parente}

\affiliation{Universidade Estadual de Roraima, Av. Senador Helio Campo, s/n - Centro, Caracaraí - RR, 69360-000}

\author{J. T. M. Pacobahyba}

\affiliation{Departamento de F\'{\i}sica, Universidade Federal de Roraima, BR 174, Km 12. Bairro Monte Cristo. CEP: 69300-000 
Boa Vista/RR}

\author{Minos A. Neto \footnote{Corresponding author.\\ E-mail address: minos@pq.cnpq.br (Minos A. Neto)\\ 
Contact number: 55-92-3305-4019}} 
\affiliation{Universidade Federal do Amazonas, Departamento de F\'{\i}sica, 3000, Japiim, 69077-000, Manaus-AM, Brazil}

\author{Ijan\'{\i}lio G. Ara\'{u}jo}

\affiliation{Departamento de F\'{\i}sica, Universidade Federal de Roraima, CEP: 69300-000, Boa Vista/RR.}

\author{J. A. Plascak}
\affiliation{Universidade Federal da Para\'{\i}ba, Centro de Ci\^encias Exatas e da Natureza - 
Campus I, Departamento de F\'{\i}sica - CCEN Cidade Universit\'aria 58051-970 - Jo\~ao Pessoa-PB, Brazil}
\affiliation{ Department of Physics and Astronomy, University of Georgia, 30602 Athens GA, USA}

\date{\today}

\begin{abstract}
\textbf{ABSTRACT}

The spin-$1/2$ anisotropic Heisenberg model with antiferromagnetic exchange interactions in the presence of a external 
magnetic field and a Dzyaloshinskii-Moriya interaction is studied by employing the usual mean-field approximation. 
The magnetic properties are obtained and it is shown that only second-order phase transitions take place for any 
values of the theoretical Hamiltonian parameters. Contrary to previous results from effective field theory, no anomalies 
have been observed at low temperatures. However, some re-entrancies still persist in some region of the phase diagram.

\end{abstract}

\maketitle

\section{Introduction\protect\nolinebreak}

The theoretical study of quantum magnetic systems has gained increased attention in the last decades not only because of their quite 
interesting intrinsic quantum phase transitions, which are induced by quantum fluctuations, but because these magnetic systems can 
also be well suited to describe some superconducting materials. For instance, the cuprate La$_2$CuO$_4$ compound has been shown to 
exhibit metamagnetic behavior \cite{thio,che} and, when doped with Sr, forming the La$_{2-x}$Sr$_x$CuO$_4$ compound, it becomes a 
high-temperature superconductor for $x>2.5\% $ \cite{bed}. 

On the theoretical point of view, the spin-$1/2$ antiferromagnetic Heisenberg model (AHM) has been one of the most studied model in 
order to try to understand the role played by the quantum fluctuations in the CuO$_2$ planes of these cuprates superconductors. The 
interest in the AHM comes from the early Anderson suggestion that these quantum fluctuations should be responsible for the superconductivity 
in this class of compounds \cite{ander}. Therefore, the AHM has been treated from several different techniques and some aspects of its 
phase diagram are well stablished (see, for instance, references \cite{minos3,minos} and references therein). However, in real materials, 
anisotropies are expected, not only regarding the exchange interactions but also from spin-orbit couplings, which may lead to the 
Dzyaloshinskii-Moriya interaction. In this sense, the AHM in the presence of an external field and a Dzyaloshinskii-Moriya 
interaction turns out to be a very interesting model with applications in several experimental realizations, going from cuprates 
superconductors to spin-glass behavior in magnetic systems \cite{minos3,minos,sun}. 

The AHM in the presence of an external field has been treated by the effective-field theory (EFT) 
\cite{minos3} and the 
phase diagram  in the temperature versus external field presents a reentrant behavior at low temperatures. 
This same model with an
external longitudinal field and an extra Dzyaloshinskii-Moriya interaction has also been studied by 
employing the EFT. The results show that 
the phase diagram also exhibits re-entrancies and, in addition, some anomaly behavior at low temperatures 
\cite{minos,minos2}. As no 
exact solution is still available for this model, the results obtained by the EFT are still questionable 
and, in some way, controversial. 

Our purpose in this work is thus to study the thermodynamic behavior of the AHM with XXZ exchange anisotropy  
in the presence of a longitudinal external magnetic field and a Dzyaloshinskii-Moriya 
interaction placed in the $z$ direction. We employ the usual mean-field approximation, based on Bogoliubov 
variational approach using a 
two-spin cluster, in order to analyze the order parameter and the corresponding phase transition. 
A two-spin like approximation has been proven to be quite efficient  in treating the transverse Ising model 
\cite{pla1} and the Blume-Capel model in a transverse crystal field \cite{pla2}.
Although still being a mean-field 
approach, it would be very interesting to compare the results so obtained in treating the AHM
with external field and Dzyaloshinskii-Moriya interactions with the previous ones from EFT.

The plan of the paper is as follows. In the next section, the model and the variational procedure formalism are presented. The results are 
discussed in Section \ref{results} and some final remarks are commented in the last section.

\section{Model and Formalism}
\label{model}

The model studied in this work is the anisotropic antiferromagnetic Heisenberg model in a magnetic field and with a Dzyaloshinskii-Moriya 
interaction, which can be described by the following Hamiltonian
\begin{eqnarray}
\mathcal{H}=J\sum_{\left\langle i,j \right\rangle}\left[\left(1-\Delta\right)\left(S_{i}^{x}S_{j}^{x}+S_{i}^{y}S_{j}^{y}\right)
+S_{i}^{z}S_{j}^{z}\right] 
%\nonumber
%\\
-\sum_{\left\langle i,j \right\rangle}\textbf{D}\cdot\left(\textbf{S}_{i}\times \textbf{S}_{j}\right)
-\sum_{i=1}^N\textbf{H}\cdot \textbf{S}_{i},
\label{hv}
\end{eqnarray}
where the first term represents the nearest-neighbors anisotropic exchange interaction, with $\Delta$ being the anisotropy parameter, the 
second term represents the nearest-neighbors antisymmetric Dzyaloshinskii-Moriya (DM) interaction placed along the $z$ direction
($\textbf{D}=\textbf{D}_{ij}=-\textbf{D}_{ji}=D{\bf \hat z}$), the third term corresponds to the Zeeman interaction with an external magnetic 
field $\textbf{H}=H{\bf \hat z}$, and finally $S^{\nu}_{i}(\nu=x,y,z)$ are the spin-$1/2$ Pauli matrices at $i$-sites in a hypercubic lattice 
of $N$ spins. On such a lattice, one has coordination number $c$ given by $c=2, ~4, ~6, \cdots$ for, respectively, the one-dimensional lattice, 
square two-dimensional lattice, simple cubic three-dimensional lattice, and so on. This model has no exact solution and the topology of the 
phase diagram in the $H-D$ plane, as a function of the anisotropy $\Delta$, still remains a fundamental 
problem. For $\Delta=0$ one has the 
isotropic Heisenberg model, while for $\Delta=1$ we recover the spin-$1/2$ Ising model, both with DM interaction.

We will employ a variational method based on Bogoliubov inequality (an inequality that is mainly based on arguments of convexity, as can be 
seen, for instance, in reference \cite{falk1970}), which can be formally written, for any classical or quantum system, as
\begin{equation}
F\left(\mathcal{H}\right)\leq F_{0}\left(\mathcal{H}_{0}\right)+\left\langle\mathcal{H}-\mathcal{H}_{0}\right\rangle_0\equiv\Phi(\eta), 
\label{b}
\end{equation}
where $F$ and $F_{0}$ are the free energies associated with two systems defined by the Hamiltonians $\mathcal{H}$ and $\mathcal{H}_{0}(\eta)$, 
respectively, the thermal average $\left\langle...\right\rangle_0$ should be taken in relation to the canonical distribution associated 
with the trial Hamiltonian $\mathcal{H}_{0}(\eta)$, with $\eta$ standing for variational parameters. The approximated free energy $F$ is given 
by the minimum of $\Phi(\eta)$ with respect to $\eta$, i.e. $F\equiv\Phi_{min}(\eta)$.

With $\textbf{D}$ and $\textbf{H}$ defined above, the hamiltonian $\mathcal H$ in eq. (\ref{hv}) can be rewritten as
\begin{eqnarray}
\mathcal{H}=J\sum_{\left\langle i,j \right\rangle}\left[\left(1-\Delta\right)\left(S_{i}^{x}S_{j}^{x}+S_{i}^{y}S_{j}^{y}\right)
+S_{i}^{z}S_{j}^{z}\right] 
%\nonumber
%\\
-D\sum_{\left\langle i,j \right\rangle}\left({S}_{i}^x{S}_{j}^y-{S}_{i}^y{S}_{j}^x\right)
-H\sum_{i=1}^N {S}_{i}^z.
\label{h}
\end{eqnarray}

Otherwise, for the trial Hamiltonian $\mathcal{H}_0$, we have chosen the simplest cluster for this model, which corresponds to a sum of $N/2$ 
disconnected pairs of spins. As we are dealing with an antiferromagnetic system, each spin of the pair belongs to one particular 
$A$ or $B$ sub-lattice. We then have
\begin{eqnarray}
\mathcal{H}_{0}=\sum_{\ell=1}^{N/2}\left\lbrace J\left[\left(1-\Delta\right)\left(S_{A_{\ell}}^{x}S_{B_{\ell}}^{x}+S_{A_{\ell}}^{y}S_{B_{\ell}}^{y}\right)+S_{A_{\ell}}^{z}S_{B_{\ell}}^{z}\right]
-D\left(S_{A_{\ell}}^{x}S_{B_{\ell}}^{y}-S_{A_{\ell}}^{y}S_{B_{\ell}}^{x}\right) 
\right. && \nonumber
\\
-\left.\left(\eta_A+H\right)S_{A_{\ell}}^{z}+\left(\eta_B+H\right)S_{B_{\ell}}^{z}\right\rbrace,
\label{h0}
\end{eqnarray}
where $S_{A_{\ell}}^{\alpha}$ and  $S_{B_{\ell}}^{\alpha}$ are the spins of the $\ell$-th pair on the $A$ and $B$ sub-lattices, respectively, 
and $\eta_A$ and $\eta_B$ are variational parameters, which are different in each sub-lattice.

It is not difficult to diagonalize the above trial Hamiltonian and to obtain the corresponding free-energy $F_0$. The same holds for the mean 
value $\left\langle\mathcal{H}-\mathcal{H}_{0}\right\rangle_0$, where we still have $Nc/2-N/2$ remaining pairs in $\mathcal H$. Thus, after 
minimizing the right hand side of Eq. (\ref{b}) with respect to the variational parameters $\eta_A$ and $\eta_B$ we obtain the approximated 
mean-field Helmholtz free energy per spin, $f=\Phi/N$, which can be written as
\begin{eqnarray}
f=-\frac{1}{2\beta}\ln\left\lbrace2e^{-K}\cosh\Delta_{1}+2e^{K}\cosh2K\Delta_{2}\right\rbrace-\frac{J}{2}(c-1)m_{A}m_{B},
\label{f}
\end{eqnarray}
with
\begin{equation}
\Delta_{1}=2h-(c-1)(m_{A}+m_{B})K,
\label{d1}
\end{equation}
\begin{equation}
\Delta_{2}=\sqrt{(1-\Delta)^{2}+(c-1)^{2}(m_{A}-m_{B})^{2}+d^{2}}
\label{d2}
\end{equation}
and the corresponding sub-lattice magnetizations $m_{A}$ and $m_{B}$ given by
\begin{equation}
m_{A}=\frac{\sinh\Delta_{1}+e^{2K}\Delta_{3}\sinh K\Delta_{2}}{\cosh\Delta_{1}+e^{2K}\cosh K\Delta_{2}}
\label{ma}
\end{equation}
and
\begin{equation}
m_{B}=\frac{\sinh\Delta_{1}-e^{2K}\Delta_{3}\sinh K\Delta_{2}}{\cosh\Delta_{1}+e^{2K}\cosh K\Delta_{2}},
\label{mb}
\end{equation}
where
\begin{equation}
\Delta_{3}=(c-1)(m_{A}-m_{B})/\Delta_{2}.
\label{d3}
\end{equation}
In the above expressions, we have defined the reduced quantities $K=J/k_{B}T$, $h=H/J$, $d=D/J$, and $\beta=1/k_BT$, where $k_B$ is the 
Boltzmann constant.

In order to analyze the criticality of this system, it will be more convenient to define an order 
parameter that characterizes the 
antiferromagnetic phase transition. The two new quantities that we will use here are the total 
($m$) and staggered ($m_{s}$) magnetizations 
(the latter one being the desired order parameter of the antiferromagnetic transition), which 
are given by
\begin{equation}
m=\frac{1}{2}(m_{A}+m_{B})
\label{7}
\end{equation}%
and 
\begin{equation}
m_{s}=\frac{1}{2}(m_{A}-m_{B}).
\label{8} 
\end{equation}
Substituting Eqs. (\ref{ma}) and (\ref{mb}) in (\ref{7}) and (\ref{8}) we have 
\begin{equation}
m=\frac{\sinh\tilde{\Delta}_{1}}{\cosh\tilde{\Delta}_{1}+e^{2K}\cosh 2K\tilde{\Delta}_{2}}
\label{m}
\end{equation}
and
\begin{equation}
m_{s}=\frac{\tilde{\Delta_{3}}e^{2K}\sinh2K\tilde{\Delta}_{2}}{\cosh\tilde{\Delta}_{1}+e^{2K}\cosh 2K\tilde{\Delta}_{2}},
\label{ms}
\end{equation}
where now 
\begin{equation}
\tilde{\Delta_{1}}=2K[h-(c-1)m],
\label{d1t}
\end{equation}
\begin{equation}
\tilde{\Delta}_{2}=\sqrt{(1-\Delta)^{2}+(c-1)^{2}m_{s}^{2}+d^{2}},
\label{d2t}
\end{equation}
and
\begin{equation}
\tilde{\Delta}_{3}=(c-1)m_{s}/\tilde{\Delta}_{2}.
\label{d3t}
\end{equation}
Accordingly, the free energy defined in (\ref{f}) can be written in terms of the new order parameter as 
\begin{equation}
f(m,m_{s})=-\frac{1}{2\beta}\ln\left\lbrace2e^{-K}\cosh\tilde{\Delta}_{1}+
2e^{K}\cosh2K\tilde{\Delta}_{2}\right\rbrace-\frac{J}{2}(c-1)(m^{2}-m^{2}_{s}).
\label{fm}
\end{equation}

Thus, for a given value of the set of Hamiltonian parameters, namely the reduced exchange 
anisotropy $\Delta/J$, the reduced 
Dzyaloshinskii-Moriya interaction $d$, and the reduced external field $h$, all in units of 
the exchange interaction $J$, we can obtain 
the temperature dependence of $m$ and $ms$ by simultaneously solving the two nonlinear 
coupled equations (\ref{m}) and (\ref{ms}). Below 
the antiferromagnetic phase transition the magnetizations of sub-lattices A and B are 
opposite and nonzero, while above the transition 
temperature one has $m_{s}=0$ and $m = m_{0}$, which is the paramagnetic phase. In this 
paramagnetic phase one has
\begin{equation}
m_{0}=\frac{\sinh(2h-2(c-1)Km)}{\cosh(2L-2(c-1)Km)+e^{2K}\cosh2K\sqrt{(1-\Delta)^{2}+d^{2}}}
\label{12}
\end{equation}
and 
\begin{eqnarray}
f(m_{0})=-\frac{1}{\beta2}\ln\left\lbrace2e^{-K}\cosh(2h-2(c-1)Km)+2e^{K}\cosh2K\sqrt{(1-\Delta)^{2}+d^{2}}\right\rbrace && \nonumber
\\ 
-\frac{J}{2}(c-1)m^{2}_{0}.
\label{13}
\end{eqnarray}

Close to the second-order transition line, one has $m_s<<1$ and the corresponding free 
energy (\ref{fm}) can be expanded in a Landau type so 
we can obtain a closed form equation for the N\'eel transition temperature $T_N$ 
(in fact $k_BT_N/J$) as a function of the Hamiltonian 
parameters. We can also obtain the criticality by noting that at the transition $m_s = 0$ 
and from relations (\ref{m}) - (\ref{d3t}) we obtain two coupled equations
\begin{eqnarray} 
m=\frac{\sinh\tilde{\Delta}_{1}}{\cosh\tilde{\Delta}_{1}+e^{2K}\cosh 2K\tilde{\Delta}_{4}},
\label{tn1}
\end{eqnarray}%
and 
\begin{eqnarray} 
1=\left(\frac{c-1}{\tilde{\Delta}_{4}}\right)\frac{e^{2K}\sinh2K\tilde{\Delta}_{4}}{\cosh\tilde{\Delta}_{1}+e^{2K}\cosh 2K\tilde{\Delta}_{4}}, 
\label{tn2}
\end{eqnarray}%
where $\tilde{\Delta}_{4}=\sqrt{(1-\Delta)^{2}+d^{2}}$. Thus, given the values of $\Delta$, 
$d$, $h$, and $c$, these two equations furnishes $m$ and $k_BT_{N}/J$.

In order to seek for first-order transitions, we have to compare the antiferromagnetic free 
energy with $m_s\ne 0$ and the paramagnetic 
one with $m_s=0$. The first-order phase transition is located when they have the same value. 
However, for the present model, as discussed in the next section, no first-order 
transitions have been detected.

\section{Results}
\label{results}

\begin{figure}[htbp]
\centering
\includegraphics[width=7.5cm,height=8.0cm]{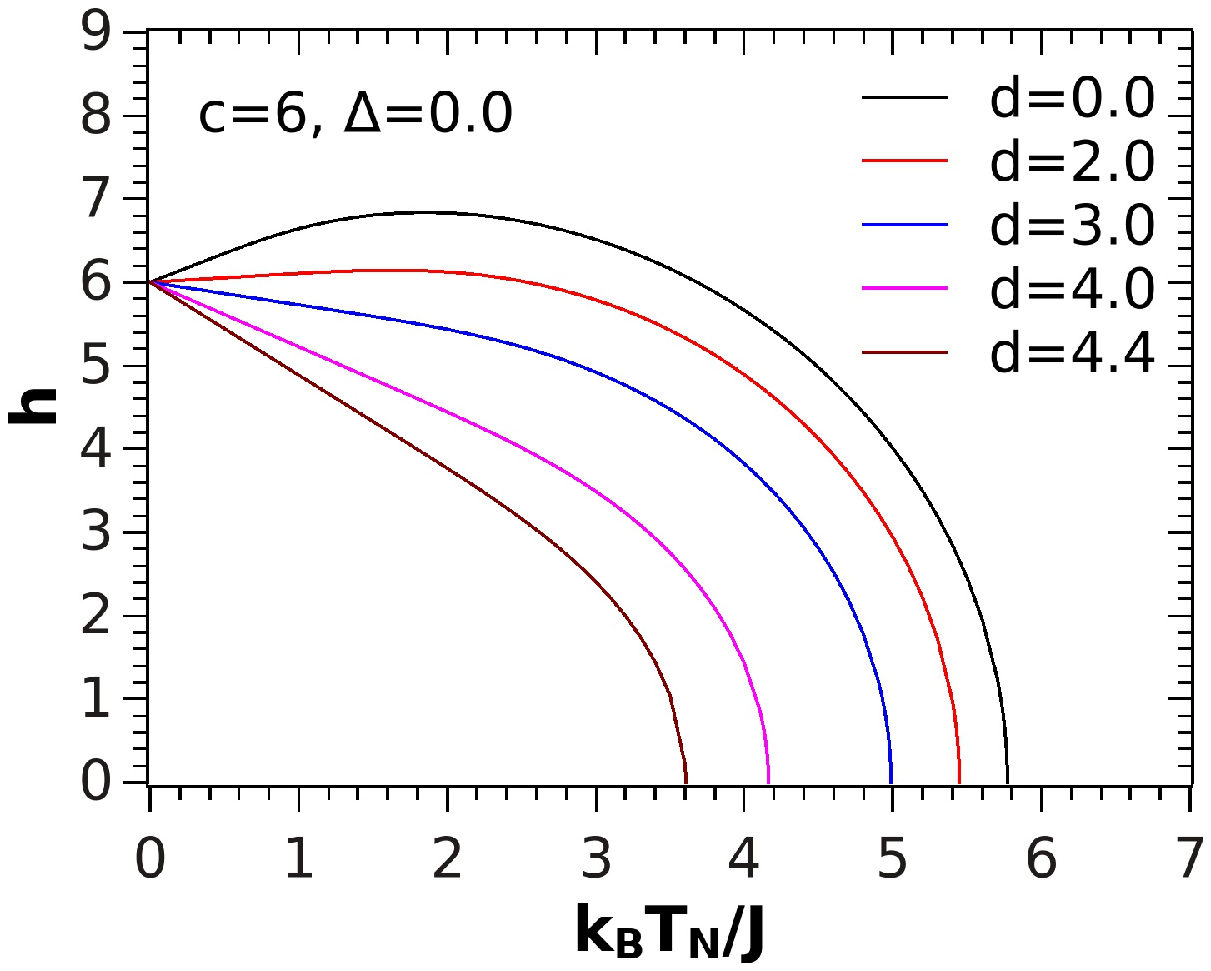}
\hspace{0.5cm}
\includegraphics[width=7.5cm,height=8.0cm]{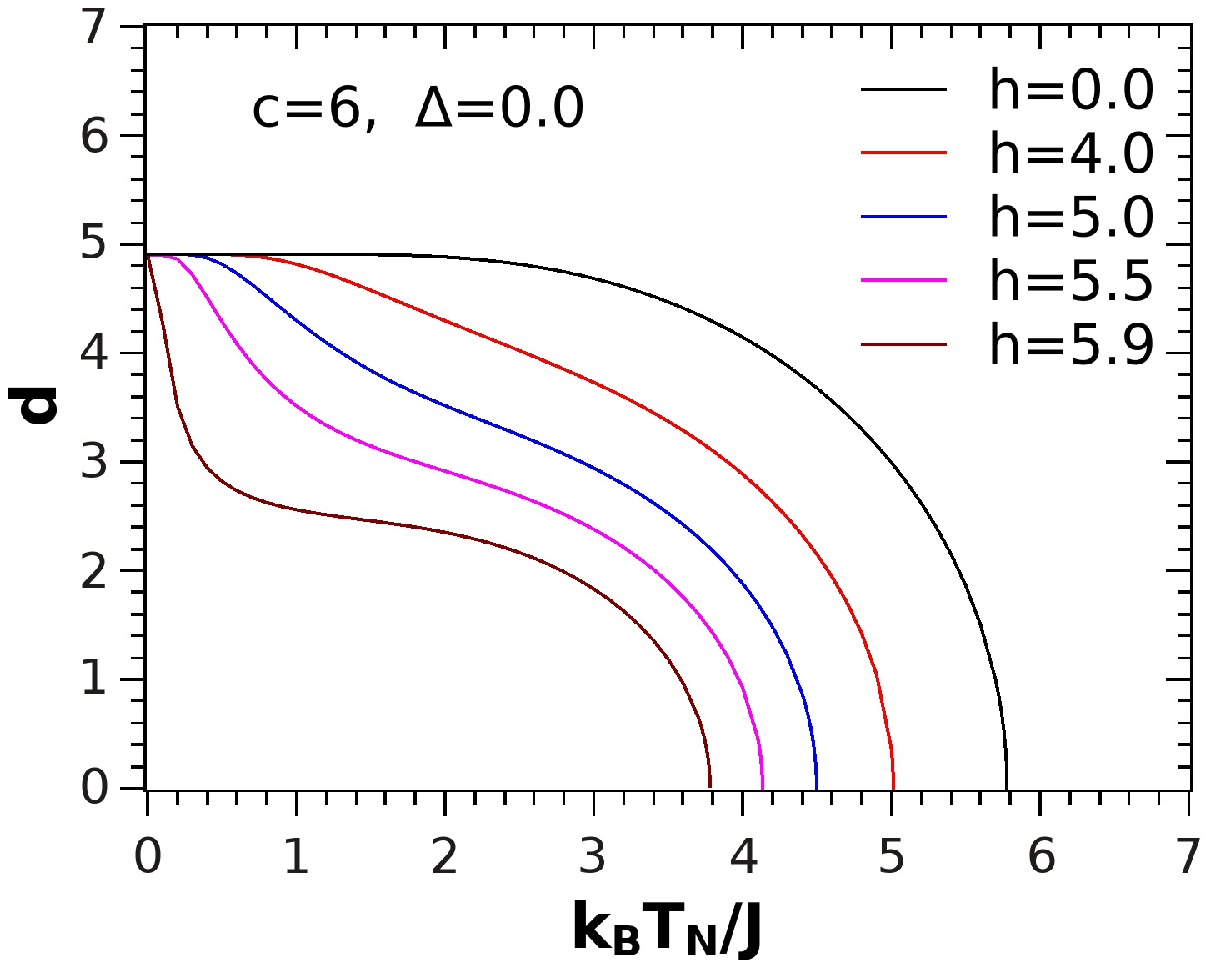}
\caption{Critical transition temperature $k_BT_N/J$ of the simple cubic lattice $c=6$ and isotropic Heisenberg model $\Delta=0$ as a 
function of: (left) external field $h$ and several values of the DM interaction $d$; and (right) DM interaction $d$ and several values 
of the external field $h$.} 
\label{c6}
\end{figure}

The reduced critical transition temperature $k_BT_N/J$, as a function of the Hamiltonian parameters, is shown in Fig. \ref{c6} for 
the simple cubic lattice and the isotropic Heisenberg model ($\Delta=0$) in the presence of external field and DM interaction. All 
the transition lines are second order and we have no indication of any first-order transition lines in these diagrams. We can see 
that the increase of either the external field or the DM interaction tends to decrease the corresponding transition temperature. 

We can see that phase diagram in the temperature versus external field presents a reentrancy 
for DM interactions $d\lesssim2.5$ while no re-entrancies are found in the phase diagram in the
temperature versus DM interaction for any value of $h$. In addition, at $T=0$, the value of the
critical external field is given by $h_c=c$ and is independent of the DM interaction. On the other
hand, from Eqs. (\ref{tn1}) and (\ref{tn2}) it can be shown
that the critical value of the DM interaction $d_c$ is given by
\begin{equation}
d_c=\sqrt{(c-1)^2-(1-\Delta)},
\end{equation}
and is independent of the external field.

The critical temperature at zero external field, namely $k_BT_{N}/J(H=0) = 5.78$, is also comparable 
to those obtained from different methods, such as EFRG-12 $k_BT_{N}/J = 4.09$ obtained by 
de Sousa and Araujo \cite{araujo}, $k_BT_{N}/J = 3.54$ from Monte Carlo simulations
\cite{he1993}, high-temperature expansion $k_BT_{N}/J=  3.59$ \cite{rush1967}, variational 
cumulant expansion $k_BT_{N}/J =4.59$ \cite{li1995}, EFT-2 $k_BT_{N}/J = 4.95$ and EFT-4 
$k_BT_{N}/J= 4.81$ \cite{minos3}.

\begin{figure}[htbp]
\centering
\includegraphics[width=7.5cm,height=8.0cm]{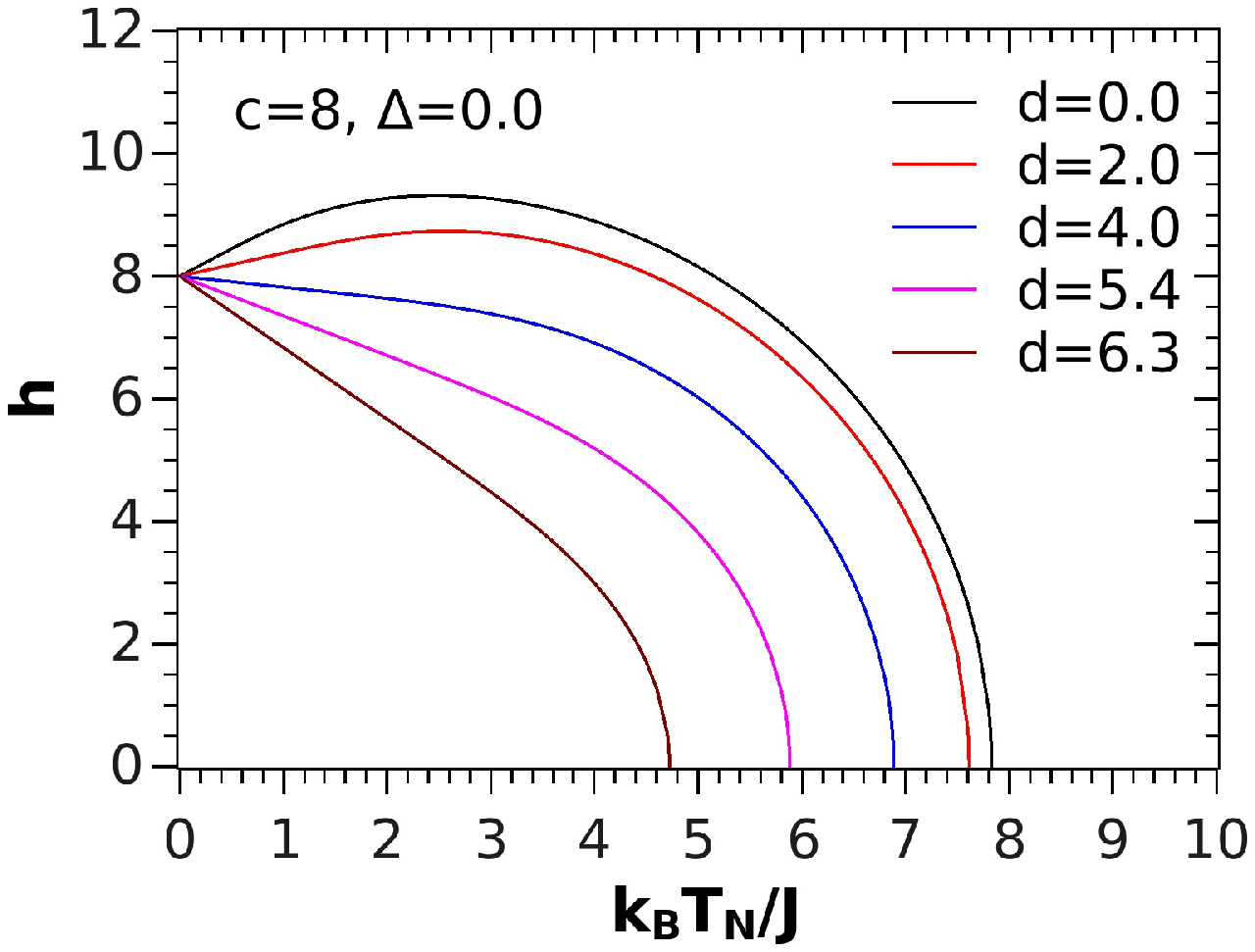}
\hspace{0.5cm}
\includegraphics[width=7.5cm,height=8.0cm]{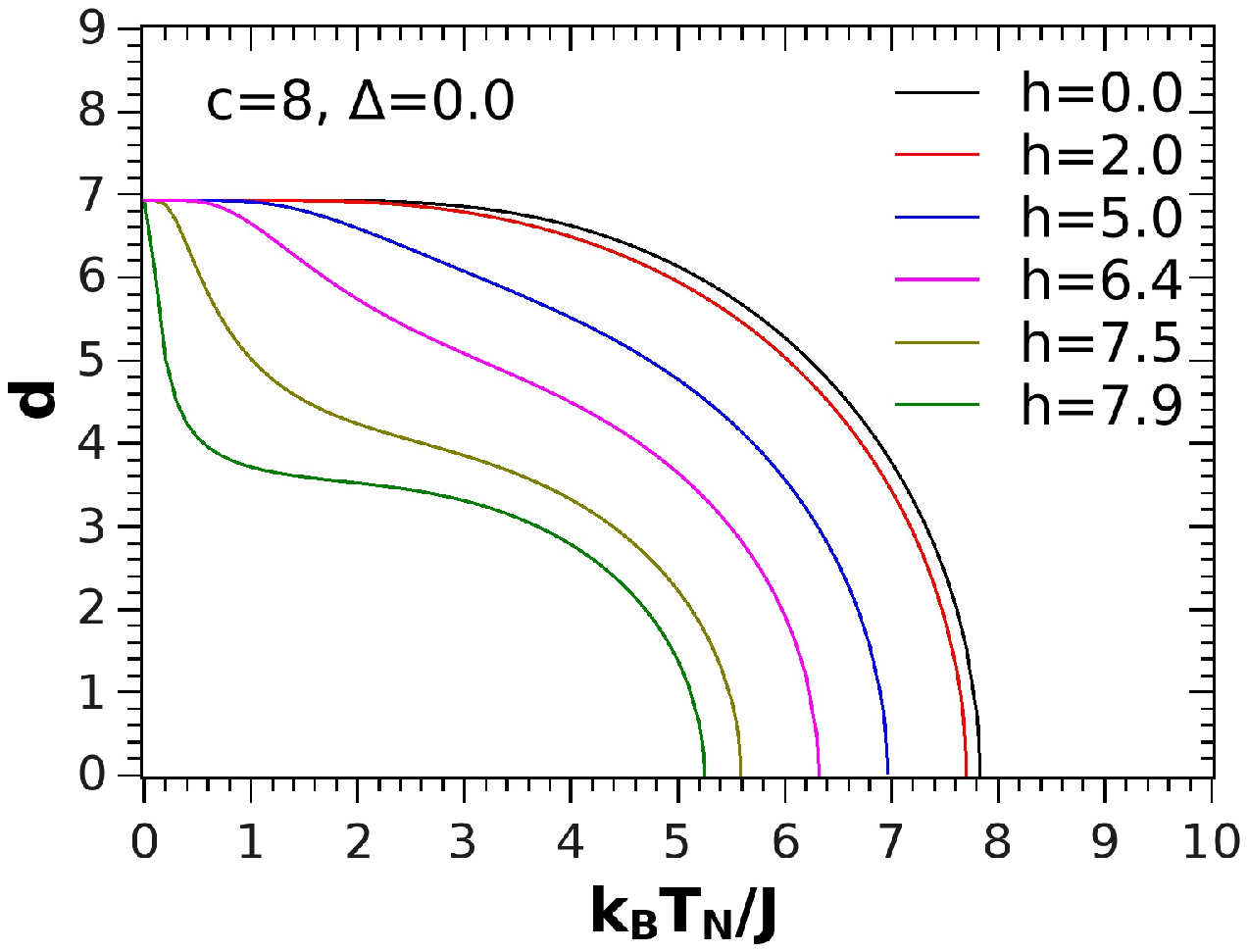}
\caption{The same as Fig. \ref{c6} for $c=8$.} 
\label{c8}
\end{figure}

The same sort of picture can be seen in Fig. \ref{c8} for higher dimensions, which is an 
example for the case $c=8$. For this value of the coordination number,
at zero field, our results coincide with those by Bublitz \textit{et al.} MFA-2 $k_BT_{N}/J 
= 7.83$ \cite{bublitzbcc}. This numerical 
result is also comparable to other different methods, such as high-temperature expansion 
$k_BT_{N}/J = 5.53$ \cite{pan}, 
EFT-2 $k_BT_{N}/J= 6.94$ , and EFT-4 $k_BT_{N}/J= 6.89$ \cite{minos3}. 

For the particular case $\Delta = d = h = 0$ we have the Ising limit. In this 
way $m = 0$ from Eq. (\ref{tn1}), as expected, and the Eq. (\ref{tn2}) provides exactly 
the equation of the pair approximation for Ising 
model \cite{tucker1998} which is given by
\begin{equation}
2K(c-1)e^{2K}=1+e^{2K}.
\label{pa}
\end{equation}

As a final comment, it should be stressed that the qualitative behavior of the phase 
diagrams are also the same for different values of the anisotropy $\Delta$, even in the 
limit of the Ising case $\Delta=1$ \cite{minos2}.
%In Fig. (\ref{mfa_diagrams_3}), we show the logarithmic behavior of $h$ and $d$ as a function of the coordination number $c$, for any value 
%of $d^{*}\leqslant d_{c}$ and $h^{*}\leqslant h_{c}$, respectively. In this figure, the anisotropy value varies from $0$ to $1$. We can 
%take the limit $T\rightarrow0$, or $K\rightarrow \infty$. In the left figure, we show the logarithmic behavior of h as function c. We 
%verified that this behavior is independent of the anisotropy value for any values of $c$ with the temperature tending to zero.

%\begin{figure}[htbp]
%\centering
%\includegraphics[width=7.5cm,height=8.0cm]{lnh_vs_z_t0_mfa.eps}
%\hspace{0.5cm}
%\includegraphics[width=7.5cm,height=8.0cm]{lnd_vs_z_t0_mfa.eps}
%\caption{the logarithmic behavior of $h$ and $d$ as a function of the coordination number $c$.} 
%\label{mfa_diagrams_3}
%\end{figure}

%For the case of Fig. (\ref{mfa_diagrams_3}) on the right, a slightly different behavior. We observed that the anisotropy effect is relevant 
%for some values of the coordination number $c$. However, when $c\rightarrow\infty$, this logarithmic behavior is no longer depend on $\Delta$. 
%At this limit, for certain conditions, we can arrive at an equation that gives the value of $d\approx\sqrt{(1-\Delta)^{2}+(c-1)^{2}}$ as a 
%function of $c$ and $\Delta$ (note that it does not depend on $h$, as seen in the figure on the right). For $c$ tending to infinity, 
%$d\varpropto c$.

\section{Final Remarks}

We have studied the anisotropic antiferromagnetic Heisenberg model, in a external magnetic 
field and with a Dzyaloshinskii-Moriya 
interaction, using a mean-field procedure based on Bogoliubov inequality for the free 
energy by employing a cluster of two spins. 
We have obtained the phase diagram as a function of the Hamiltonian parameters. 

When comparing our results to those obtained from the effective field theory, which are,
up to our knowlwdge, the only results for this system, we notice that there is no anomalous 
behavior of the 
transition lines at low temperature, although some re-entrancies are still present in some
region of the phase diagram.

Of course, this treatment is still a mean field one. It is in some sense different 
from the EFT, although the latter one having also an 
intrinsic mean field character. The present results strongly suggest that the behavior of 
the thermodynamic properties of the anisotropic 
Heisenberg model with Dzyaloshinskii-Moriya interaction is far from having a complete 
understandable behavior. 

\textbf{ACKNOWLEDGEMENTS}

This work was partially supported by FAPEAM and CNPq (Brazilian Research Agencies).


\begin{thebibliography}{1}

\bibitem{thio} T. Thio, et al., Phys. Rev. B \textbf{38} 905 (1988).

\bibitem{che} S. W. Cheong, et al., Phys. Rev. B \textbf{39} 4395 (1989).

\bibitem{bed} J. G. Bednorz, K.A. Muller, Z. Phys. B \textbf{64} 89 (1986).

\bibitem{ander} P. W. Anderson, Science \textbf{235} 1196 (1987).

\bibitem{minos3} Minos A. Neto, J. Roberto Viana, J. Ricardo de Sousa, J. Magn. Magn. Mater. \textbf{324} 2405 (2012). 

\bibitem{minos} Walter E. F. Parente, J. T. M. Pacobahyba, Ijan\'ilio G. Ara\'ujo, Minos A. Neto, J. Ricardo de Sousa, \"Umit Akinci 
J. Magn. Magn. Mater. \textbf{355} 235 (2014).

\bibitem{sun} Yunzhou Sun, Lin Yi, Xige Zhao, Huiping Liu, Solid State Commun. \textbf{144} 61 (2007).

\bibitem{minos2} Walter E. F. Parente, J. T. M. Pacobahyba, Ijan\'i­lio G. Ara\'ujo, Minos A. Neto, J. Ricardo de Sousa,  
Physica E \textbf{74} 287 (2015).

\bibitem{pla1} J. A. Plascak, Phys. Stat. Sol. B {\bf 120} 215 (1983).

\bibitem{pla2} D. C. Carvalho and J. A. Plascak, Physica A {\bf 432} 240 (2015).

\bibitem{falk1970} H. Falk, Am. J. Phys. \textbf{38}, 858 (1970).

\bibitem{araujo} J. Ricardo de Sousa, I. G. Araujo, J. Magn. Magn. Mater \textbf{202}, 231 (1999).

\bibitem{he1993} He-Ping Ying \textit{et al.}, Phys. Lett. A \textbf{183}, 441 (1993).

\bibitem{rush1967} G. S. Rushbrooke and P. J. Wood, Mod. Phys. \textbf{11}, 409 (1967).

\bibitem{li1995} Hong Li and T. L. Cheng, Phys. Rev. B \textbf{52}, 15979 (1995).

\bibitem{bublitzbcc} E. Bublitz Filho, J. Ricardo de Sousa, J. Magn. Magn. Mater. \textbf{269}, 266 (2004).

\bibitem{pan} K. K. Pan, Phys. Lett. A \textbf{244}, 169 (1998).

\bibitem{tucker1998} J. W. Tucker, T. Balcerzak, M. Gzik, A. Sukiennicki, J. Magn. Magn. Mater. \textbf{187} (1998) 381.


\end{thebibliography}
\end{document}